\documentclass[a4paper, 10 pt]{article}

\usepackage{amsmath}
\usepackage{graphicx}
\usepackage{epsfig}
\usepackage{subfig}
\usepackage{xcolor}
\usepackage{amsfonts}
\usepackage{cite}
\usepackage{ntheorem}
\usepackage{amssymb}

\usepackage{enumitem}

\theorembodyfont{\normalfont}
\theoremseparator{:}
\theoremheaderfont{\bfseries}
\newtheorem*{prf}{Proof}
\theoremseparator{.}
\theoremheaderfont{\itshape\bfseries}
\newtheorem*{thm}{Theorem}
\newtheorem{rmk}{Remark}
\newtheorem{asm}{Assumption}

\newtheorem{prop}{Proposition}

\newcommand{\cmnt}[1]{\ignorespaces}
\newcommand{\ie}{\textit{i}.\textit{e}. }
\newcommand{\eg}{\textit{e}.\textit{g}. }
\newcommand{\Poincare}{Poincar\'{e} }
\newcommand{\PoincareBendixson}{Poincar\'{e}-Bendixson }

\newcommand{\ds}{\dot{s}}
\newcommand{\dds}{\ddot{s}}
\newcommand{\df}{f^\prime}
\newcommand{\ddf}{f^{\prime \prime}}
\newcommand{\ps}{s^\prime}

\title{\LARGE \bf
A Data Driven Vector Field Oscillator with Arbitrary Limit Cycle Shape
}

\author{Venus Pasandi\textsuperscript{\em a,b}, Aiko Dinale\textsuperscript{\em b}, Mehdi Keshmiri\textsuperscript{\em a}, Daniele Pucci\textsuperscript{\em b}
	\thanks{\textsuperscript{\em a}Isfahan University of Technology, Isfahan, Iran,\newline
		{\tt venus.pasandi@me.iut.ac.ir, mahdik@cc.iut.ac.ir}
	}%
	\thanks{\textsuperscript{\em b}Istituto Italiano di Tecnologia, Genoa, Italy, \newline
		{\tt aiko.dinale@iit.it, daniele.pucci@iit.it}}%
}

\begin{document}

\maketitle
\thispagestyle{empty}
\pagestyle{empty}

\begin{abstract}
Cyclic motions in vertebrates, including heart beating, breathing and walking, are derived by a network of biological oscillators having fascinating features such as entrainment, environment adaptation and robustness.
These features encouraged engineers to use oscillators for generating cyclic motions.
To this end, it is crucial to have oscillators capable of characterizing any periodic signal via a stable limit cycle.
In this paper, we propose a 2-dimensional oscillator whose limit cycle can be matched to any periodic signal depicting a non-self-intersecting curve in the state space.
In particular, the proposed oscillator is designed as an autonomous vector field directed toward the desired limit cycle.
To this purpose, the desired reference signal is parameterized with respect to a state-dependent phase variable, then the oscillator's states track the parameterized signal.
We also present a state transformation technique to bound the oscillator's output and its first time derivative.
The soundness of the proposed oscillator has been verified by carrying out a few simulations.
\end{abstract}

\section{INTRODUCTION}

Nonlinear oscillators have been widely used by the engineering community to model and control physical phenomena \cite{strogatz2018nonlinear,zhao2016cpg,yu2014survey}.
Their interesting features such as entrainment, synchronization and smooth modulation of the output signal make them appropriate for robotics applications such as cyclic motions of manipulators or legged robot locomotion.
Using oscillators for generating the reference trajectory or control signal provides capabilities of smoothness, continuity, disturbance rejection and adaptation.
The oscillator encodes the desired reference trajectory or control signal via a stable limit cycle.
The essence of well-known oscillators, like the Matsuoka's and Hopf's, is their capability of generating limit cycles with a specific shape. 
However, a specific limit cycle shape constraints the types of signals that can be generated.
In this paper, we propose a two dimensional oscillator which can generate any periodic trajectory, depicting a non-self-intersecting curve in the state space.

Assuming the desired limit cycle is defined by a Lyapunov function, the problem of designing a dynamical system with a desired limit cycle is expressed as the problem of constructing a dynamical system for a desired Lyapunov function \cite{hirai_method_1972,green1984synthesis}. Besides controlling the limit cycle, such framework has been extended to non-autonomous dynamical systems to design transient trajectories and achieve the desired convergence \cite{ohno_synthesis_2006}. Although these algorithms are interesting from the mathematics point of view, they can not be directly applied for engineering purposes because of their lack of analytical predictability. In fact, these algorithms generate different dynamic structures for different desired limit cycles and the properties of the dynamics, like domain of attraction and attracting rate, are not determined a priori.

A widespread strategy for designing a dynamical system generating an arbitrary periodic signal is to transform a well-understood dynamical system into an oscillating one with desired limit cycle. For instance, a linear spring-damper system can generate a variety of cyclic signals with the help of a forcing term. 
To design an autonomous system, the forcing term is defined by a nonlinear function of a phase variable and learned by standard machine learning techniques~\cite{ijspeert_dynamical_2013}.
From a more general perspective, the limit cycle of a phase oscillator is mapped to the desired periodic trajectory in the state space through a phase-dependent scaling function.
Thus, a general family of nonlinear phase oscillators which can track almost any continuous trajectory is constructed~\cite{ajallooeian_general_2013}.
The phase of the dynamical systems proposed in \cite{ijspeert_dynamical_2013} and \cite{ajallooeian_general_2013} is generated by an independent phase dynamics which results in trajectory tracking and not limit cycle tracking.
Furthermore, the desired trajectory is asymptotically stable and not asymptotically orbitally stable.
To provide limit cycle tracking of a desired periodic trajectory, a Hopf's oscillator is altered by two nonlinear functions which are determined such that the \PoincareBendixson theorem is satisfied in a predefined neighborhood of the limit cycle~\cite{ajallooeian_design_2012}.
The framework guarantees local stability and it is not straightforward to extend it for achieving global stability, since the \PoincareBendixson theorem is no longer applicable.

Another method for designing an oscillator is to use a data driven vector field which has been originally proposed for generating a discrete system with an arbitrary limit cycle~\cite{hirai_synthesis_1982}. The discrete vector field generated in the neighborhood of the limit cycle is approximated by a function, like a polynomial one, and hence, a continuous dynamical system with a locally stable desired limit cycle is created~\cite{okada_polynomial_2002}. 
However, to the best of the authors' knowledge, the possibility of designing a continuous nonlinear vector field for ensuring the global stability of the desired limit cycle has not been explored yet.

In the present paper, we propose a two dimensional continuous dynamical system that can track any non-self-intersecting closed trajectory in the state space. The main idea is to generate a data driven vector field directed toward the desired trajectory. To this purpose, the desired trajectory is parameterized with respect to a state-dependent phase variable. Then, the oscillator dynamics is designed to track the parameterized trajectory. Moreover, we propose a state transformation method for generating a bounded output by using the proposed oscillator. 
The contribution of this paper is threefold. 
First, the proposed system provides asymptotic orbital stability of the desired trajectory.
Second, the convergence to the desired trajectory is irrespective of the parameters of the system.
Third, the proposed oscillator is capable of generating bounded output.

The rest of the paper is organized as follows. Section \ref{sec_bkg} introduces the notations and definitions used in the paper. It also recalls the concepts of orbital stability and transverse dynamics which will be used to prove the asymptotic orbital stability of the desired trajectory in the proposed oscillator. Section \ref{sec_mdldev} presents the model development of the proposed oscillator. 
In Section \ref{sec_addfet}, we modify the proposed oscillator to bound the output and its first time derivative. 
Section~\ref{sec_vld} reports simulation results. Finally, Section \ref{sec_cnl} concludes the paper with a few remarks and perspectives.
  
\section{BACKGROUND}\label{sec_bkg}

\subsection{Notations and Definitions}
\begin{itemize}[leftmargin=4.0mm]
\item $\mathbb{R}$ and $\mathbb{R}^+$ are the set of real and positive real numbers.
\item The $i^{th}$ component of a vector $\mathbf{q} \in \mathbb{R}^m $ is written as $q_i$.
\item  Given a function $g(x(t)):\mathbb{R} \rightarrow \mathbb{R}$ where $t$ represents the time, its first derivatives with respect to $x$ and $t$ are denoted as $g^\prime= \frac{dg}{dx}$ and $\dot{g} = \frac{dg}{dt}$, respectively.
\item  A \emph{$ \mathcal{C}^k $-function} is a function with $k$ continuous derivatives.
\item  A function ${f(t):[0,\infty) \rightarrow \mathbb{R}}$ is a \emph{$T$-periodic function} if for some positive constant $p$, we have ${f(t+p) = f(t)}$ and $T$ is the smallest $p$ with such property.
\item  A \emph{simple closed curve} is a continuous closed curve that does not cross itself. In mathematical word, $\gamma : [a,b] \rightarrow \mathbb{R}^n$ is a simple closed curve if $\gamma (a) = \gamma(b)$ and additionally $ \gamma(c) \neq \gamma (d)$, $\forall c,d \in [a,b)$. Hence, the simple closed curve $\gamma$ is a one-to-one mapping from $[a,b)$ to $\mathbb{R}^n$.	
\end{itemize}	
	
\subsection{Stability of a periodic trajectory}
Given the dynamical system $\dot{\mathbf{x}} = f(\mathbf{x})$, where $\mathbf{x} \in \mathbb{R}^n$ is the state vector, the periodic trajectory $\mathbf{x}^*(t)$ is
\begin{itemize}[leftmargin=4.0mm]
\item \emph{stable} if $\forall \epsilon>0, \, \exists \delta>0$ such that
\begin{equation*}\label{eq_stablityDef}
\|\mathbf{x}(t_0)-\mathbf{x}^*(t_0)\| < \delta \quad  \Rightarrow \quad \|\mathbf{x}(t)-\mathbf{x}^*(t)\| < \epsilon,
\end{equation*}
\item \emph{asymptotically stable} (AS) if it is stable and $\exists \eta>0 $ that
\begin{equation*}\label{eq_ASDef}
\|\mathbf{x}(t_0)-\mathbf{x}^*(t_0)\| < \eta \quad  \Rightarrow \quad \lim\limits_{t \rightarrow \infty} \mathbf{x}(t)=\mathbf{x}^*(t),
\end{equation*}
\item \emph{orbitally stable} (OS) if $\forall \epsilon>0, \, \exists \delta>0$ such that
\begin{equation*}\label{eq_OSDef}
\inf_\varphi\|\mathbf{x}(t_0)-\mathbf{x}^*(\varphi)\| < \delta \, \Rightarrow \, \inf_\varphi\|\mathbf{x}(t)-\mathbf{x}^*(\varphi)\| < \epsilon,
\end{equation*}
\item \emph{asymptotically orbitally stable} (AOS), called also stable limit cycle, if it is orbitally stable and $\exists \eta>0$ that
\begin{equation*}\label{eq_AOSDef}
\inf_\varphi\|\mathbf{x}(t_0)-\mathbf{x}^*(\varphi)\| < \eta \, \Rightarrow \, \lim\limits_{t\rightarrow \infty} \inf_\varphi\|\mathbf{x}(t)-\mathbf{x}^*(\varphi)\| = 0.
\end{equation*}
\end{itemize}

The above stability definitions are stated from \cite{hale1971functional}.
It is noteworthy that to verify either the OS or the AOS, we consider the time evolution of the distance between the system's states and the closed set of the trajectory $\mathbf{x}^*$.
On the other hand, to prove the stability or the AS, we examine the time evolution of the distance between the system's states and a specific point of $\mathbf{x}^*$ which changes with respect to time. Therefore, the stable/AS conditions are stricter than~OS/AOS.

\subsection{Transverse Dynamics}
The OS property of a periodic trajectory is usually investigated through two techniques.
In the first technique, the stability of a periodic trajectory of a continuous system is attributed to the stability of the equilibrium point of the corresponding discrete map, called \Poincare map.
A \Poincare map, known also as \emph{first return map}, is the intersection of the system's trajectory in the state space with a \Poincare section, that is a lower-dimensional hypersurface transversal to the trajectory under study.
Usually, it is not possible to find the \Poincare map analytically.
Therefore, a linearization of the \Poincare map is often computed numerically, and its eigenvalues are used to verify local OS.
In the second technique, the limit cycle is considered as a class of invariant sets and its stability is investigated through the LaSalle's invariance principle.
For this purpose, one uses a Lyapunov function that equals to zero along the trajectory and is strictly positive elsewhere.
An approach for constructing such Lyapunov function is the \emph{transverse dynamics}, commonly called also \emph{moving \Poincare sections}.
In this approach, a transversal hypersurface is defined in the state space which moves along the trajectory under study.
Considering an \textit{n}-dimensional $T$-periodic trajectory $\mathbf{x}^*$, a hypersurface $\sigma(\varphi)$ is defined for $\varphi \in [0,T]$ where $\sigma(\varphi)$ is transversal to $\mathbf{x}^*(\varphi)$, \ie $ \dot{\mathbf{x}}^*(\varphi) \notin \sigma(\varphi) $ and $\sigma(0) = \sigma(T)$.
Then, a new coordinate system $(e,\phi)$ is established where the scalar $\phi$ represents which of the transversal surfaces $\sigma$ is inhibited by the current state $\mathbf{x}$, and the \emph{transverse coordinate} $e \in \mathbb{R}^{n-1}$ determines the location of $\mathbf{x}$ within the hypersurface $\sigma(\phi)$, with $e=0$ implying that $\mathbf{x} = \mathbf{x}^*(\phi)$.
The dynamics of the transverse coordinate is the transverse dynamics.
The stability of the equilibrium point $e = 0$ of the transverse dynamics ensures the OS of the trajectory $\mathbf{x}^*$.
In this way, one can analyze the stability analytically and characterize the stability region.
For more information, the reader can refer to~\cite{leonov2006generalization,shiriaev2008can,tang2014transverse}.

\section{DATA DRIVEN VECTOR FIELD OSCILLATOR}\label{sec_mdldev}
In this section, we design a continuous 2-dimensional dynamical system that provides asymptotic orbital stability of any desired $T$-periodic function $f(t):[0,\infty) \rightarrow \mathbb{R}$ depicting a simple closed curve in the state space. From hereafter, we consider one period of $f(t)$, \ie we assume the $T$-periodic function $f(t)$ as $f(t):[0,T) \rightarrow \mathbb{R}$.

Consider the 2-dimensional dynamical system described by the following differential equation
\begin{equation}\label{eq_NeuronDynamics}
\dds = \ddf(\varphi)-\alpha \left( \ds-\df(\varphi) \right)-\beta \left( s-f(\varphi) \right),
\end{equation}
where $ \mathbf{s} = \left( s,\ds \right) \in \mathbb{R}^2 $ represent the states, $ \alpha, \beta \in \mathbb{R}^+ $ are coefficients and $\varphi$ is the phase variable defined based on $\mathbf{s}$~as
\begin{equation}\label{eq_geivenOrbitParameterization}
{\small
	\varphi(\mathbf{s}) = 
	\begin{cases}
	f^{-1}(f_l) & s \leq f_l \\
	\{\varphi|f(\varphi) = s, \, \df(\varphi) \ds \geq 0 \} & f_l \leq s \leq f_u, \ds \neq 0 \\
	\{\varphi|f(\varphi) = s, \, \df(\varphi) \geq 0 \} & f_l \leq s \leq f_u, \ds = 0 \\
	f^{-1}(f_u) & s \geq f_u,
	\end{cases}}
\end{equation} 
where $f_l$ and $f_u$ are the lower and upper bounds of $f$. 

\begin{figure}[!t]
	\centering
	\includegraphics[width=3.0in]{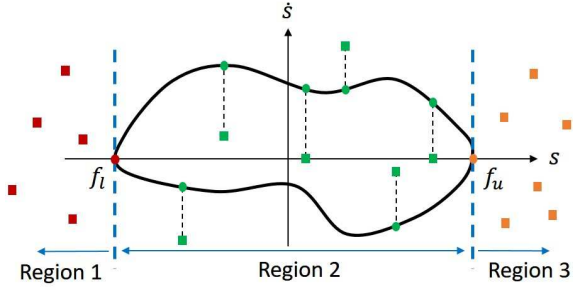}
	\setlength{\belowcaptionskip}{-2.0mm}
	\caption{Geometrical representation of the target point. For every state of the system (squares), there is a corresponding target point (circles) on the desired trajectory (black curve).}	
	\label{fig_GoalPointDetermination}
\end{figure}

Conceptually, the proposed dynamics (\ref{eq_NeuronDynamics}) follows the trajectory of the \emph{target point}, \ie a point defined on the function $f$ in the state space with coordinates $(f(\varphi(\mathbf{s})), \df(\varphi(\mathbf{s})))$.
So, the target point is associated with the states and represents a parametrization of the function $f$ with respect to $\varphi$.
Fig. \ref{fig_GoalPointDetermination} shows the relation between the states and the corresponding target points.
The black curve is the function~$f$.
Squares represent states at different time instants while circles are the corresponding target points.
Based on the phase variable definition (\ref{eq_geivenOrbitParameterization}), the state space is divided into three regions.
For all the states in Region 1 $(R_1)$, the target point is the red circle with coordinates $(f_l,0)$, while for all the states in Region 3 $(R_3)$, the target point is the orange circle with coordinates $(f_u,0)$. For the states in Region 2 $(R_2)$, the target point is a point on the function $f$ with the same $s$ coordinate as the state of the system. 
We call the dynamics (\ref{eq_NeuronDynamics}) along with the phase defined in \eqref{eq_geivenOrbitParameterization} as \textit{Data driven Vector field Oscillator} (DVO).
The remainder of this section is devoted to investigating the properties of the DVO.

\begin{rmk}
	As the function $f(t)$ is a simple closed curve in the state space, the target point assigned to $\mathbf{s}$ is unique.
\end{rmk}	 

In light of the above, we assume from now on that

\begin{asm}\label{asm_desiredOrbit}
	$f(t)$ is a $T$-periodic $ \mathcal{C}^3 $-function as
	\begin{equation}
	f(t):[0,T) \subseteq \mathbb{R} \rightarrow \mathbb{R},
	\end{equation}
	describing a simple closed orbit in the plane $(f,\dot{f})$.
\end{asm}

\begin{rmk}\label{rmk_pointTracking}
	In addition to the limit cycle tracking, one can use DVO for point tracking where the trajectory $f$ is constant. In the case of point tracking, the dynamics (\ref{eq_NeuronDynamics}) is simplified to the well-known PD control as
	\begin{equation}
	\dds = -\alpha \ds - \beta (s-f).
	\end{equation}
\end{rmk}

\begin{thm}\label{thm_stabilityDVO}
	Given that Assumption \ref{asm_desiredOrbit} is satisfied, and $\alpha$ is a positive function and $ \beta $ is a positive constant then the trajectory $f(t)$ is the semi-stable limit cycle of the DVO expressed in \eqref{eq_NeuronDynamics}-\eqref{eq_geivenOrbitParameterization}.
	The region of attraction of $f$ is
	\begin{equation}
		\mathcal{D}_o = \{ (s,\ds) \, | \, \ds^2 \geq f^{\prime 2}(\varphi) \}.
	\end{equation}
\end{thm}

The theorem states that a trajectory converges to $f$ if it is initialized outside the closed curve of $f$ in the state space.

\begin{prf}
	Let us define a weighted error $e$ as
	\begin{equation}\label{eq_weighted_error}
		e = \frac{1}{2} (\ds^{2}-f^{\prime 2}) + \dfrac{\beta}{2}\left(s-f\right)^2 - \ddf \left( s-f \right).
	\end{equation}
	If $(s,\ds)=\left( f,\df \right)$, then we have $e=0$.
	To show also that if $e=0$ then $(s,\ds)=\left( f,\df \right)$, let us specify $e$ in the three regions $R_1, R_2$ and $R_3$, defined based on the phase definition \eqref{eq_geivenOrbitParameterization}.
	In the case $\mathbf{s} \in R_2$, phase definition \eqref{eq_geivenOrbitParameterization} results in $s = f$ and the weighted error is simplified as
	\begin{equation}
		e = \dfrac{1}{2} \left( \ds ^2 - f^{\prime 2} \right).
	\end{equation}
	Thus, $\ds = \df$ if $e=0$.
	In the case $\mathbf{s} \in R_1$, we have $s<f_l$, $f = f_l$, $\df = 0$ and $\ddf > 0$.	
	Thus, $e$ is simplified~as
	\begin{equation}
	e = \frac{1}{2} \ds^{2} + \dfrac{\beta}{2}\left(s-f\right)^2 - \ddf \left( s-f \right),
	\end{equation}
	which is the sum of three positive terms. 
	Thus $(s,\ds)=\left( f,\df \right)$ if $e=0$.	
	The same results are true also if $\mathbf{s} \in R_3$, but in this case $s>f_u$, $f = f_u$, $\df = 0$ and $\ddf < 0$.
	Consequently, $e=0$ iff the states $\mathbf{s}$ coincide the trajectory of $f$ in the state space, i.e. $(s,\ds)=\left( f,\df \right)$.
	Thus, we consider $ v = \frac{1}{2}e^{2} $ as the candidate Lyapunov function for proving the semi-stability of the limit cycle $f$.
	To compute the time derivative of $v$, one needs to compute the time derivative of the phase variable $\varphi$ on which the trajectory $f$ depends.
	Based on the phase definition (\ref{eq_geivenOrbitParameterization}), $\varphi$ is continuous  for $\mathbf{s} \in \mathcal{D}_o$ and thus, the time derivative of $\varphi$ is
	\begin{equation}\label{eq_phaseDynamics}
		\dot{\varphi} =
		\begin{cases}
			\dfrac{\ds}{\df} & \mathbf{s} \in R_2\\
			0 & \mathbf{s} \in R_{1,3},
		\end{cases}
	\end{equation}
	where $ R_{1,3} = R_1 \cup R_3 $.
	Therefore, the time derivative of the weighted error along the dynamics (\ref{eq_NeuronDynamics}) is as follows
	\begin{equation}\label{eq_de}
		\dot{e} = 
		\begin{cases}
			-\alpha\ds \left( \ds-\df \right) & \mathbf{s} \in R_2 \\
			-\alpha\ds^{2} & \mathbf{s} \in R_{1,3}.
		\end{cases}
	\end{equation}
	
    The time derivative of $ v $ for $\mathbf{x} \in \mathcal{D}_o$ is obtained as
	\begin{equation}
		\dot{v} =
		\begin{cases}
			-\frac{\alpha}{2} \ds \left(\ds+\df\right) \left(\ds-\df\right)^{2} & \mathbf{s} \in R_2 \\
			-\frac{\alpha}{2} \ds^{2} \left(\ds^{2} +\beta \left(s-f\right)^{2} - 2 \ddf\left(s-f\right) \right) & \mathbf{s} \in R_{1,3},
		\end{cases}
	\end{equation} 
    which is negative semi definite because $ \ds \df \geq 0 $ for $\mathbf{s} \in R_2$ and $ \ddf(s-f) \leq 0 $ for $\mathbf{s} \in R_{1,3}$. 
    Thus, $e$ is bounded. This implies that the states $\mathbf{s}$ are bounded if the trajectory $f$ and its first derivative $\df$ are bounded.
	For asymptotic results, it is sufficient to examine the largest invariant subset of the set $ \Omega = \{ \mathbf{s} : \dot{v} = 0 \} $. Considering the dynamics (\ref{eq_NeuronDynamics}), one verifies that $ \{ e=0 \} $ is the only invariant set of $ \Omega $. 
	Therefore, asymptotic stability of $ e = 0 $ is concluded based on the LaSalle lemma.
	The proof is completed by showing the radially unbounded property of the Lyapunov function $v$ which is obvious from the definition of $v$.
	\hfill\ensuremath{\blacksquare}
\end{prf}

	\begin{rmk}
	If the set $\mathcal{D} = \mathbb{R}^2 - \mathcal{S}$ where
	\begin{equation}
		\mathcal{S} := \{ (s,\ds) \, | \, s \in (f_l,f_u), \, \ds = 0 \},
	\end{equation}
	is a positive invariant set of the DVO then $\varphi$ is also continuous in the inside of the closed curve of $f$. 
	Thus, the proof is satisfied for $\mathbf{s} \in \mathbb{R}^2$.
	Consequently, one can conclude that $f$ is the globally stable limit cycle of the DVO.
	\end{rmk}
	Assuring the positive invariancy of $\mathcal{D}$ is not possible as the matter of the continuity of the dynamical system \eqref{eq_NeuronDynamics}.
	Albeit, one can define the coefficient $\alpha$ such that $\mathcal{D}$ is almost an invariant set, \ie for $\delta > 0$, the set $\mathcal{D}_\delta = \mathbb{R}^2 - \mathcal{S}_\delta$ where
	\begin{equation}
		\mathcal{S}_\delta := \{ (s,\ds) \, | \, s \in (f_l+\delta,f_u-\delta), \ds = 0 \},
	\end{equation}
	is a positive invariant set. 
	Thus, $f$ is almost global stable limit cycle of DVO \ie the trajectory of the DVO converges to $f$ from almost any initial condition.
	The following proposition suggests a definition for $\alpha$ which results in such a small $\delta$ that the trajectories converge to $f$ from any initial condition in practice.
	
\begin{prop}
	Given that the following assumptions hold
	\begin{itemize}
		\item Assumption \ref{asm_desiredOrbit} is satisfied,
		\item $\beta$ is positive constant, and
		\item $\alpha$ is defined as
		\begin{equation}
			\alpha = \dfrac{\alpha_1}{\alpha_b} \left( \alpha_b + \tanh(f^{\prime \prime 2}) \right),
		\end{equation}
		where
		\begin{equation}
			\alpha_b = \tanh \left( \alpha_2 f^{\prime 2} + \alpha_3 \ds^2 + \alpha_4 (s-f)^2 \right) + \epsilon,
		\end{equation}
		and $\alpha_1,\alpha_2, \alpha_3, \alpha_4, \epsilon \in \mathbb{R}^+$ are constants and $\epsilon \ll 1$. 
	\end{itemize}
	then $f$ is the almost globally stable limit cycle of the DVO.
\end{prop}

\section{Considering Output Limits}\label{sec_addfet}

The proposed DVO has been mainly conceived for performing cyclic motions in robotics applications.
If we define the output of the DVO as $y(t) = s(t)$, then we can generate a cyclic signal, tracking a predefined desired trajectory, and use it as the reference signal for the robot controller or directly as the control signal.
Considering this scenario, it becomes necessary to provide the possibility of generating a bounded output to avoid physical limitations of the robot such as position, velocity or actuator limits. 
The rest of this section investigates the problem of output limits in details.

Assume that the feasible region of the output is as
\begin{equation}\label{eq_outputLimits}
	\mathcal{Q} := \{ y \in \mathbb{R} \, : \, y_{min} < y < y_{max}, \ |\dot{y}| < \delta_{\dot{y}} \},
\end{equation}
where $y_{min}, y_{max}, \delta_{\dot{y}} \in \mathbb{R}$ are constants denoting the minimum and maximum of the output $y$, and the maximum feasible magnitude of $\dot{y}$.
To preserve the feasible region (\ref{eq_outputLimits}), we introduce the following output definition
\begin{equation}\label{eq_outputDefinition}
	y = y_{avg} + \delta_y \tanh\left( s(\tau) \right),
\end{equation}
where $ y_{avg}=\frac{y_{min}+y_{max}}{2} $, $ \delta_y = \frac{y_{max}-y_{min}}{2} $ and $\tau(t)$ is an exogenous state with the following dynamics
\begin{equation}\label{eq_timeParam}
	\dot{\tau} = \frac{\delta_{\dot{y}} \tanh{(\ps)}}{J_s \ps},
\end{equation}
where $J_s = \delta_y \left( 1-\tanh^2(s) \right)$.

Given (\ref{eq_outputDefinition}) and (\ref{eq_timeParam}), the time derivative of the output $y$ is
\begin{equation}
	\dot{y} = \delta_{\dot{y}} \tanh(\ps).
\end{equation}
Consequently, the output definition (\ref{eq_outputDefinition}) guarantees that the output limits are preserved, \ie $y \in \mathcal{Q}$.

Now, we can write the DVO with respect to $\tau$ as
\begin{equation}\label{eq_boundedNeuronDyn}
	s^{\prime \prime} = g_a(\varphi)-\alpha \left( \ps-g_v(\varphi) \right)-\beta \left( s-g_p(\varphi) \right),
\end{equation}
with
\begin{equation}
\begin{aligned}
	g_p(\varphi) &= \tanh^{-1} \left( \dfrac{f(\varphi)-y_{avg}}{\delta_y} \right), \\
	g_v(\varphi) &= \tanh^{-1} \left( \dfrac{\df(\varphi)}{\delta_{\dot{y}}} \right), \\
	g_a(\varphi) &= \dfrac{\delta_y \left( 1-\tanh^2(g_p) \right)}{\delta_{\dot{y}}^2 (1-\tanh^2(g_v))} \dfrac{g_v}{\tanh(g_v)} \ddf,
\end{aligned}
\end{equation}
and
\begin{equation}\label{eq_phaseParameterizationConstrained}
	{\small
	\varphi(\mathbf{s}) =
	\begin{cases}
		g^{-1}_p(g_l) & s \leq g_l \\
		\{\varphi|g_p(\varphi) = s, \, g_v(\varphi) \ps \geq 0 \} & g_l \leq s \leq g_u, \ps \neq 0 \\
		\{\varphi|g_p(\varphi) = s, \, g_v(\varphi) \geq 0 \} & g_l \leq s \leq g_u, \ps = 0 \\
		g^{-1}_p(g_u) & s \geq g_u,
	\end{cases}}
\end{equation}
where $ g_l $ and $g_u$ are the lower and upper bounds of $g_p$.

Integrating \eqref{eq_boundedNeuronDyn}, one can compute $s(\tau)$, but we are still missing $s(t)$ which is required to compute the output $y(t)$. To overcome this problem, let us define new states $(s_1,s_2) = \left( s(t), \ps(t) \right)$ and rewrite the dynamics \eqref{eq_boundedNeuronDyn} with respect to the new states as following
\begin{equation}\label{eq_constrainedOscillator}
\begin{cases}
	\dot{s}_1 = \frac{\delta_{\dot{y}} \tanh(s_2)}{J_s } \\
	\dot{s}_2 = \frac{\delta_{\dot{y}} \tanh(s_2)}{J_s s_2} \left( g_a - \alpha(s_2 - g_v) - \beta(s_1-g_p) \right).
\end{cases}
\end{equation}
Hence, one can integrate the dynamics (\ref{eq_constrainedOscillator}) with respect to the time $t$ and calculate $y(t)$.

We call the dynamics (\ref{eq_constrainedOscillator}) expressed with respect to the phase definition (\ref{eq_phaseParameterizationConstrained}) as the Modified DVO (MDVO).
	
\section{VALIDATION}\label{sec_vld}

In this section, we illustrate the DVO and MDVO performance when tracking a desired reference signal through a few numerical simulations.

\subsection{Asymptotic Stability vs. Asymptotic Orbital Stability}
\begin{figure}[!b]
	\centering
	\includegraphics[width=3.3in]{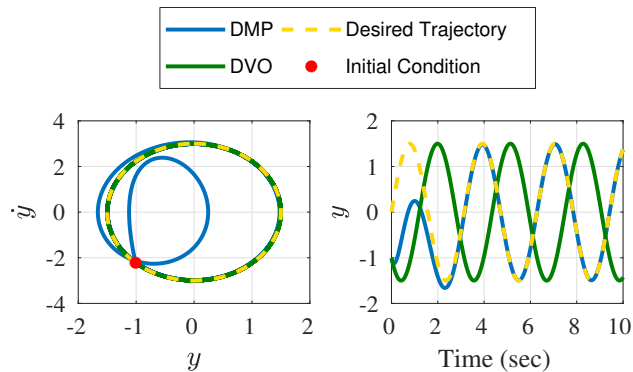}
	\setlength{\belowcaptionskip}{-2.0mm}
	\caption{The behavior of DVO vs. DMP when tracking a simple sinusoidal trajectory.}
	\label{fig_DMP}
\end{figure}
We compared AS and AOS from a mathematical point of view in Section \ref{sec_bkg}. 
Instead, to explore their difference from a practical point of view, we compared the response of the DMP, an autonomous system with AS trajectory proposed in~\cite{ijspeert_dynamical_2013}, and the DVO, as an oscillator with AOS trajectory. To this purpose, we simulated these two systems when tracking the simple sinusoidal signal $f = 1.5 \sin(2t)$ from the initial condition located on the desired trajectory.
Fig.~\ref{fig_DMP} shows the behavior of the two systems in the state space (left plot) and in the time domain (right plot).
As it can be seen in the state space plot, the DVO remains on the desired trajectory but the DMP leaves the desired trajectory as the initial condition is not equal to the desired initial value.
The time domain plot shows that the steady state response of the DMP is in-phase with the desired trajectory as the initial phase is chosen to be zero, while there is a phase difference between the steady state response of the DVO and the desired trajectory. 
In particular, the steady state phase difference of the DMP is always equal to the chosen initial phase. However, the steady state phase difference of the DVO is not constant and is related to the initial conditions and convergence rate. 
This behavior is a consequence of the fact that the desired trajectory is an invariant set in the DVO  but not in the DMP.
So, we can say that the DMP imposes a \emph{time constraint} on the system response, \ie the system states must assume the desired value at specific time instants. 
The DVO, instead, imposes a \emph{timing constraint}, \ie the system states replicate the desired trajectory while guaranteeing the desired timing.
More precisely, the system states assume the desired value but not at specific time instants.
For application such as legged robot locomotion where respecting the timing constraint is only required, an oscillator with AOS trajectory, as the DVO, is more appropriate than a system with AS trajectory in terms of tracking and control effort.

\begin{figure}[!b]
	\centering
	\includegraphics[width=3.3in]{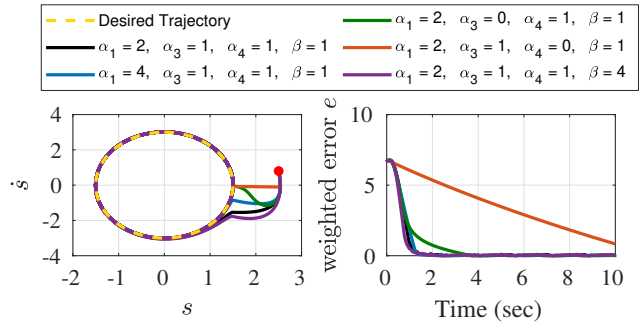}
	\caption{The effect of coefficients $(\alpha_1,\alpha_3,\alpha_4,\beta)$ in the DVO structure. The coefficient $\alpha_2 = 1$ is constant.}
	\label{fig_basicNeuronParamAnalysis}
\end{figure}

\subsection{The Effect of the Coefficients in DVO Structure}
\begin{figure}[!t]
	\centering
	\includegraphics[width=3.3in]{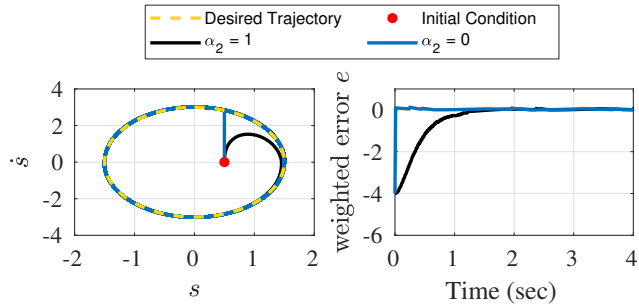}
	\setlength{\belowcaptionskip}{-2.0mm}
	\caption{The effect of coefficients $\alpha_2$ in DVO. All the remaining coefficients are constants ($\alpha_1=2, \alpha_3, \alpha_4, \beta = 1$).}
	\label{fig_basicNeuronParamAnalysis2}
\end{figure}

To analyze the effect of different coefficients in the DVO structure, let us define two quantities: \emph{reaching phase} and \emph{reaching time}.
The first one is the difference between the phase of the point at which the trajectory reaches the limit cycle and the initial phase, while the second one is the time required to reach the limit cycle. 
Fig.~\ref{fig_basicNeuronParamAnalysis} depicts the DVO response when tracking the sinusoidal signal $f = 1.5 \sin(2t)$ for five different values of the coefficients $(\alpha_1,\alpha_3,\alpha_4,\beta)$ which mostly affect the motion in $R_{1,3}$. As $\alpha_1$ increases and $\beta$ decreases (\eg the blue and purple trajectories in Fig. \ref{fig_basicNeuronParamAnalysis}), the reaching phase decreases. Though, as the time plot of the weighted error in Fig.~\ref{fig_basicNeuronParamAnalysis} shows, these coefficients do not affect much the reaching time. For $\alpha_3 = 0$, the system has a high damping coefficient when $|s-f|$ and $|f^\prime|$ are small. Similarly, for $\alpha_4=0$, the damping coefficient is high when $|\ds|$ and $|f^\prime|$ are small. 
In this way, the system converges to the limit cycle with small velocity and acceleration (\eg the green and brown trajectories in Fig.\ref{fig_basicNeuronParamAnalysis}) which results in high reaching time. 
As Fig. \ref{fig_basicNeuronParamAnalysis2} illustrates, the coefficient $\alpha_2$ influences the DVO behavior when the system is in $R_2$ and $|\ds|$ is small. In this case, the coefficient $\alpha$ increases and thus, the system experiences high acceleration which is unnecessary and also~undesirable.

\subsection{MDVO Performance}
\begin{figure}[!b]
	\centering
	\includegraphics[width=3.4in]{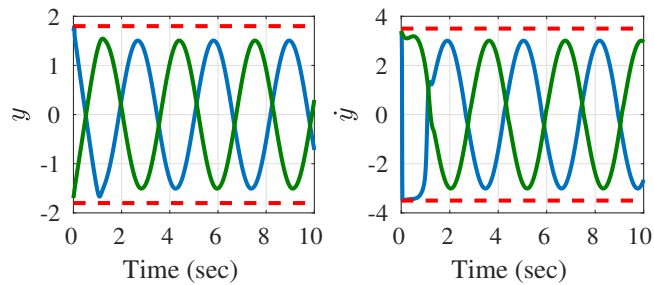}
	\caption{MDVO response for two different initial conditions. The red dashed lines are the output limits. The coefficients are chosen as $\alpha_1 = \beta = 1$, $\alpha_2 = \alpha_3 = \alpha_4 = 2$.}
	\label{fig_outputLimits}
\end{figure}

Given the sinusoidal signal $f = 1.5 \sin(2t)$ as input, we simulated the MDVO with output limits $|y|<1.8$ and $|\dot{y}|< 3.5$ for two different initial conditions $(y_0,\dot{y}_0) = (\pm1.7,3.4)$.
As can be seen in Fig. \ref{fig_outputLimits}, the output limits are preserved. 
If the chosen initial conditions are close to the both upper limits of the output, then there is a spike in the second derivative of the output at the beginning of the motion (\eg the blue trajectory).
Note that the two trajectories, with different initial conditions, do not necessarily converge together because the MDVO provides limit cycle tracking not trajectory tracking.

\subsection{Changing the Desired Trajectory}
\begin{figure}[!t]
	\centering
	\includegraphics[width=3.0in]{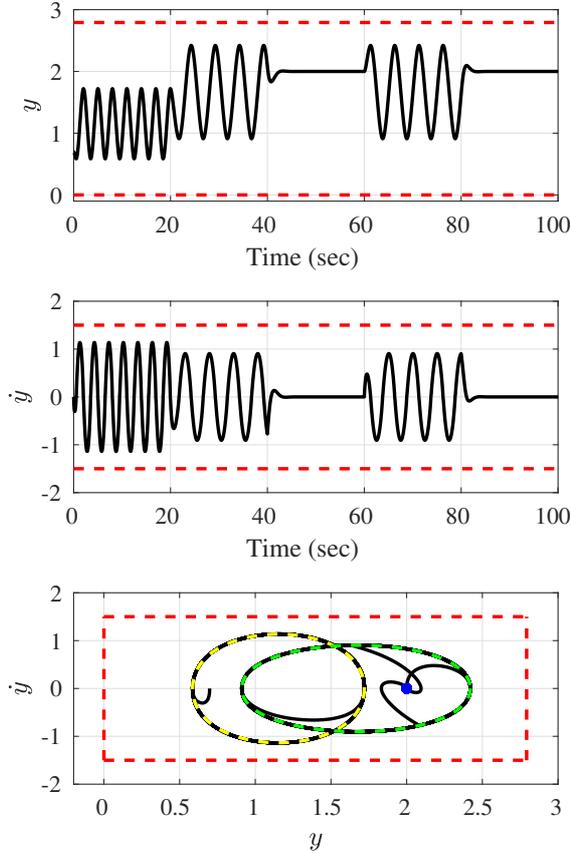}
	\setlength{\belowcaptionskip}{-2.0mm}
	\caption{MDVO's performances when changing the desired trajectory. The red dash lines are the output limits, yellow and green dashed curves are the desired trajectories $f_1$ and $f_2$, and the blue point is the desired constant trajectory $f_3$. The coefficients are $\alpha_1 = \alpha_2 = \alpha_3 = \alpha_4 = \beta = 2$. The output limits are defined as $0<y<2.8$ and $|\dot{y}|< \frac{\pi}{2}$.}
	\label{fig_changeMotion}
\end{figure}

The response of the MDVO when changing the desired motion between three functions $f_1$, $f_2$ and $f_3$ is depicted in Fig.~\ref{fig_changeMotion}. In particular, $f_1$ and $f_2$  are two periodic functions with different amplitudes and frequencies, and $f_3$ is a constant function. The desired trajectory is changed every $20$ seconds while the coefficients of the oscillator are kept constant during the simulation. As can be seen, the output is smooth and its first time derivative $\dot{y}$ is continuous.
As the MDVO is a second order differential equation, $\dot{s}_2$ and, consequently, the second time derivative of the output are not continuous.

\section{CONCLUSION}\label{sec_cnl}
We presented a novel oscillator specifically designed for those robotic applications where it is required to perform cyclic motions.
The proposed oscillator is named DVO and it is a continuous 2-dimensional dynamical system which can converge to any periodic trajectory depicting a non-self-intersecting curve in the state space.
Compared with existing results, our approach provides global asymptotic orbital stability of the periodic function and, the stability property is irrespective of the parameters of the system.
In addition, the proposed dynamical system can be used for tracking both periodic and constant functions.
This property becomes important for those applications where both periodic motions and constant posture are required.
Using the proposed dynamics, one can also generate a smooth modulation when switching from one desired trajectory to another.
Moreover, we proposed a modified version of the DVO, named MDVO, where we introduced a parameterization technique for satisfying the predefined limits on the output signal and its first time derivative.
All the above mentioned properties have been validated through simulations.

The proposed dynamical system generates a one dimensional output, and so it can control only one degree of freedom of a robotic system.
This means that for a robot with $n$ degrees of freedom, we will need $n$ DVOs (or MDVOs), \ie one for each degree of freedom.
Thus, it becomes crucial to be able to synchronize multiple systems of such kind to generate a multi-dimensional output.
In our future work, we will propose a technique to construct a synchronous networks of DVOs or MDVOs.

\bibliographystyle{unsrt}
\bibliography{ANonlinearOscillatorWithArbitraryLimitCycleShape}

\end{document}